\documentclass{article}
\usepackage{spconf,amsmath,graphicx}
\usepackage[table]{xcolor}

\title{Deducing the severity of psychiatric symptoms from the human voice}
\name{Rita Singh$^1$, Justin Baker$^2$, Luciana Pennant$^2$, Louis-Philippe Morency$^3$ \thanks{Dr. Singh was supported by the U.S. Department of Homeland Security under Award Number 2009-ST-061-CCI002-07.}}
\address{$^1$Computer Science Department, Carnegie Mellon University, Qatar \\
  $^2$McLean Hospital, Harvard Medical School, USA \\
	$^3$Language Technologies Institute, Carnegie Mellon University, USA \\
  {\small \tt rsingh@cs.cmu.edu, jtbaker@partners.org, lpennant@mclean.harvard.edu,  morency@cs.cmu.edu}
}

\begin{document}
\ninept
\maketitle
\begin{abstract}
Psychiatric illnesses are often associated with multiple
 symptoms, whose severity must be graded for accurate
diagnosis and treatment. This grading is usually done by trained
clinicians based on human observations and judgments made within doctor-patient sessions.
Current research provides sufficient reason to expect that the
human voice may carry biomarkers or signatures of many, if not all,
these symptoms.  Based on this conjecture, we explore the possibility of
objectively and automatically grading the symptoms of psychiatric
illnesses with reference to various standard psychiatric rating
scales. Using acoustic data from several clinician-patient interviews
within hospital settings, we use non-parametric models to learn and
predict the relations between symptom-ratings and voice. In the
process, we show that \textit{different} articulatory-phonetic units of speech
are able to capture the effects of \textit{different} symptoms \textit{differently}, and
use this to establish a plausible methodology that could be employed for
automatically grading psychiatric symptoms for clinical purposes.
\end{abstract}


\noindent{\bf Index Terms}: Deducing psychological states, human voice, speech, grading psychiatric symptoms, psychometry

\section{Introduction} \label{sec:intro}
Mental illnesses are characterized by a set of symptoms which manifest to different degrees of severity in different illnesses. The term \textit{grading} (or rating) of any symptom refers to the determination of the degree of severity of the symptom as expressed on a numerical scale. Currently, such grading is done manually by trained clinicians in hospital settings. This paper addresses the problem of automatically grading the symptoms of mental illnesses in patients through the analysis of their \textsl{voice}. 

\textbf{Prior studies:} It is well known that psychiatric illnesses affect different aspects of the human voice.
There is a vast body of literature in multiple fields such as behavioral ecology,  experimental psychology, social psychology, developmental psychology,  neurology and psychiatry, behavioral sciences, cognitive neuroscience, physiology etc. that report such studies. For instance, human voice has been correlated to symptoms of mental disorders such as schizophrenia \cite{a22}, depression \cite{a81, a95}, autism \cite{a35}, Parkinson’s disease \cite{a72, a77}, Huntington's disease \cite{a96}, suicidal tendencies, etc. A wide range of personal and socially relevant traits that are also significant in psychiatric assessments are correlated with features of human voice, such as emotional state \cite{a100}, dominance and attractiveness \cite{b6}, threat potential \cite{b30}, social status \cite{b34}, personality \cite{b18}, sexual orientation, level of self-consciousness etc. Voice has also been correlated with a number of other medically relevant factors that are known to indirectly affect a person's psychological state, such as the presence of diseases \cite{a72, a77}, hormone levels \cite{a82}, use of prescription medication \cite{a97}, etc.

However, most of these studies address the problem of \textit{detection} of a psychiatric disorder or disease, i.e, given a population of normal and psychiatrically ill people, these studies use voice to detect the presence or absence of various illnesses in the population. Some studies have used voice-derived cues for the \textit{differentiation} of different conditions or diseases of the brain from psychiatric disorders, e.g. \cite{flint1992acoustic} which explores the acoustic cues in the differentiation of Parkinson's disease from major depression. 

Our goal is different. We address the specific problem of \textit{grading} the symptoms of people who are already known to have particular mental illnesses. 
Specifically, our objective is to identify articulatory-phonetic cues of psychiatric illnesses through the analysis of the compositional units of speech.
Our approach is predicated on the fact that speech is produced by the movements of the articulators in the vocal tract, which have been shown to be influenced by multiple biophysical factors including a person's mental and physical state. 
Studies in multiple fields such as clinical linguistics, phonetics, phoniatrics \cite{b5,b9,a85} etc. clearly indicate that there is a strong connection between articulator movements and a speaker's physical and mental state at the time of speaking. 
Based on these studies, our hypothesis is that different psychiatric symptoms will affect articulators and articulation of speech differently, and hence manifest in different components of speech. We also hypothesize that each of these symptoms will be manifested in {\em some} articulatory units, but not others. As a consequence, any standard aggregate analysis of speech to measure or grade these symptoms may result in the dilution of the evidence from the units that exhibit the effects of these symptoms, by noise from other units that do not correlate to them. In this work, we attempt to identify the specific articulatory-phonetic units of speech that best predict the grades of different symptoms of psychiatric illnesses in patients, so that these can be used successfully for automatic grading. We do this through statistical analysis of the predictions of these grades, where the predictors are trained with acoustic features derived from these phonetic units. Our experiments are conducted on a collection of manually-rated doctor-patient recordings. The psychiatric rating scales used for assigning the manual ratings are described in detail in Section \ref{sec:strategy}.

We find that several symptoms cannot be predicted from articulatory-phonetic cues. Of those than \textit{can} be predicted, some exhibit strong correlations to specific phonetic units, but not others, while others are more generally exhibited across all sounds. Not only does this analysis give us key information on what aspects of speech to focus on to grade psychiatric illnesses, it also makes it possible to further hypothesize the effect of these illnesses on specific components of the speech production mechanism which may potentially assist in diagnoses and further understanding of the physical manifestations of these psychiatric illnesses.

\section{Articulatory phonetic connections to the speaker's mental state} \label{sec:speech}
Human voice is produced when the vocal tract induces a transformation of aerodynamic energy to acoustic energy. Several articulators play a key role in this, such as the teeth, lips, hard palate,  soft palate (velum), alveolar ridge, tongue (front, back or middle/sides, i.e. apex, dorsum and laminus respectively), uvula, glottis, pharynx etc. In its basic form, and devoid of intelligible content, the sound produced in the vocal tract is an \textit{excitation} signal that resonates within the chambers of the vocal tract. This \textit{excitation} signal is further modulated into the sound patterns characteristic of speech by the physical movements of the articulators in the vocal tract. The movements change the shape and dimensions of the various resonant chambers of the vocal tract, causing time-varying resonance patterns in the acoustic signal. This sequence of resonance patterns in the acoustic signal is perceived as (often) intelligible speech by the listener. These resonance patterns characterize \textit{phonemes}, which are considered to be the compositional units of speech from an articulatory-phonetic standpoint. 

From an articulatory-phonetic perspective, phonemes are further divided into several categories and subcategories. Each of these categories is characterized by a canonical configuration of the articulators and the vocal tract. Each phoneme thus has a \textit{locus}, which is an ideal configuration of the vocal tract necessary for its clear and complete enunciation by the speaker. In continuous speech, as one phoneme leads into another, the vocal tract changes shape continuously, moving from the locus of one phoneme to another. In reality, the loci are never completely achieved. In fact, the degree to which the articulators achieve these canonical configurations is highly characteristic of the speaker's natural healthy (or otherwise) movement abilities. Many studies that seek to identify a speaker's characteristics from voice take advantage of this fact, e.g \cite{allen2003individual, singhicassp2016}.

In addition to this, we have evidence that articulator movements are affected by many factors that relate to the physical and mental state of the speaker \cite{farnetani1997coarticulation, macneilage1970motor, guenther2006cortical, grigos2009changes}. Many biophysical factors exert considerable influence on the human speech-production mechanism, including but not limited to  physiological factors such as age, anthropometric factors such as body size, height, weight etc., a plethora of transient conditions such as intoxication, different diseases, psychological health etc. All of these factors can affect human voice by influencing parameters such as the  size, tension and agility of the vocal cords, the length of the vocal tract, the power and resonance of the voice source, {\em i.e.} the lungs, the size and shape of the resonant cavities, muscle response in the vocal apparatus etc. The effect of different psychological illnesses on muscle agility and facial muscle movements has been demonstrated in many studies, and these studies do not exclude the involvement of muscles that move the articulators in the vocal tract. The vocal-tract biology is in fact highly likely to be affected by any factor that affects the biology of the rest of the body, and the effects are likely to manifest in the quality of the speech signal produced. 

Following this line of reasoning, we expect that different categories of phonemes, which involve different sets of articulators in their production, are likely to be affected differently by a speaker's mental state. We also expect that different phoneme categories are likely to reflect the degree or severity of each symptom differently. Our experiments are designed to explore the veracity of these statements. We use the same phoneme categorizations as described in \cite{singhmipro2016}, wherein we have used an articulatory-phonetic approach for forensic applications. We describe these categorizations briefly below.
 
\subsection{Phoneme categorizations}
In the English language, phonemes are divided into two broad categories -- vowels and consonants, and two other categories that involve some intersections of these -- semivowels and diphthongs. A full description of the articulatory-phonetic bases for these divisions is given in \cite{singhmipro2016}. For our purpose, Fig. \ref{fig:consonantschart} shows the key subcategories of consonants, narrowed to those typical of North American English only. Note that we do not study vowels  for now, since these are more easily influenced by many factors such as accent, physical state etc. of the speaker, and any correlation that we deduce with mental symptoms may be heavily influenced or masked by these biases. On the other hand, past studies indicate that the consonants are less likely to be influenced \cite{cutler2000constraints} by these issues, although not negligibly so.


Consonants are categorized based on voicing, manner of articulation, and articulator placement. Voicing refers to the activity of the vocal cords: a consonant is said to be \textit{voiced }if the vocal cords vibrate during its production, and \textit{unvoiced} if they do not. The categorization of the manner of articulation refers to the formation of vocal tract configurations that involve  a) stoppage and release of airflow (\textit{plosives}), b) turbulent airstream generation (\textit{fricatives}), c) airflow obstruction with turbulent release (\textit{affricates}), d) airflow obstruction and release through nasal cavity (\textit{nasals}), e) airflow around the tongue  (\textit{liquids}), f) narrowed airflow between palate and tongue (\textit{glides}). Finally, from an articulator placement perspective, based on location of the key articulators that play a role in their production, the phonemes subcategories we consider are: a) both lips (\textit{bilabial}), b) lips and teeth (\textit{labiodental}), c) upper and lower teeth (\textit{interdental}), d) alveolar ridge (\textit{alveolar}), e) hard and soft palate (\textit{palatal}), f) velum (\textit{velar}) and glottis (\textit{glottal}).

\section{Rating of psychiatric symptoms} \label{sec:strategy}
Psychiatric illnesses are often accompanied by symptoms that can be subjectively evaluated by clinicians. The degree to which each symptom is exhibited by an individual is indicated quantitatively by rating it on one of several psychiatric rating scales that are widely accepted by the medical community. 

\noindent \textbf{The Brief Psychiatric Rating Scale (BPRS)} is one such scale  \cite{overall1962brief}. BPRS comprises a set of 24 symptoms, all or a subset of which may be scored in any specific psychiatric evaluation. The ratings for each symptom range from 1-7, and the standard practice is to assign integer values within this range. 
The original set of symptoms included under the BPRS scale are shown in Table \ref{tab:allscales}. Since its inception, modified versions of BPRS have been proposed and used, e.g. \cite{lukoff1986manual}. 

\noindent \textbf{The Montgomery-Asberg Depression Rating Scale (MADRS)} comprises a set of ten diagnostic symptoms that quantify the severity of depressive episodes in patients with mood disorders. The rating is assigned in recommended scale-steps: 0,2,4 and 6 or between these: 1,3,5. Higher scores indicate increasing severity of depressive symptoms. 

\noindent \textbf{The Positive and Negative Syndrome Scale (PANSS)} comprises 32 symptoms. These are rated 1-7 (1:absent, 2:minimal, 3:mild, 4:moderate, 5:moderate-severe, 6:severe, 7:extreme). 

Symptoms included under the BPRS, MADRS and PANSS scales are shown in Table \ref{tab:allscales}.
In our experiments, we use an additional set of two cumulative ratings listed as P01 and P02 under the PANSS scale. 
\begin{table*}[ht]
\begin{tabular}{|p{5.5cm}|p{5.5cm}|p{5.5cm}|}
\hline
\cellcolor{gray!10} \textbf{PANSS} & 4 G7 Motor retardation  & 3 B13 Self-neglect \\
44 P01 General scale total & 3 G8 Uncooperativeness & 2 B14 Disorientation \\
21 P02 Positive scale total & 6 G9 Unusual thought content & 4 B15 Conceptual disorganization   \textemdash\\
3 P1 Delusions & 3 G10 Disorientation  \textemdash & 3 B16 Blunted affect \\
4 P2 Conceptual disorganization  \textemdash & 4 G11 Poor attention  \textemdash & 4 B17 Emotional withdrawal   \textemdash\\
6 P3 Hallucinatory behavior & 2 G12 Lack of judgment and insight & 3 B18 Motor retardation   \textemdash \\
6 P4 Excitement & 3 G13 Disturbance of volition & 4 B19 Tension   \textemdash\\
7 P5 Grandiosity  \textemdash & 2 G14 Poor impulse control  & 4 B20 Uncooperativeness  \\
5 P6 Suspiciousness/persecution & 4 G15 Preoccupation  \textemdash & 6 B21 Excitement  \\
5 P7 Hostility & 4 G16 Active social avoidance  \textemdash & 4 B22 Distractability   \textemdash\\
5 N1 Blunted affect  \textemdash & \cellcolor{gray!10} \textbf{BPRS} &  6 B23 Motor hyperactivity \\
5 N2 Emotional withdrawal  \textemdash & 2 B1 Somatic concerns  \textemdash & 3 B24 Mannerisms and posturing  \textemdash \\
3 N3 Poor rapport & 5 B2 Anxiety   \textemdash &  \cellcolor{gray!10} \textbf{MADRS} \\
4 N4 Passive/apathetic social withdrawal & 6 B3 Depression  &  6 M1 Apparent sadness  \\
3 N5 Difficulty in abstract thinking  \textemdash & 5 B4 Suicidality  & 5 M2 Reported sadness  \\
4 N6 Lack of spontaneity/conversation flow & 5 B5 Guilt   & 4 M3 Inner tension  \\
5 N7 Stereotyped thinking  \textemdash & 3 B6 Hostility   & 4 M4 Reduced sleep  \\
3 G1 Somatic concern  \textemdash & 6 B7 Elated Mood   & 2 M5 Reduced appetite   \textemdash\\
5 G2 Anxiety  \textemdash & 6 B8 Grandiosity   & 4 M6 Concentration difficulties   \textemdash \\
5 G3 Guilt feelings & 5 B9 Suspiciousness   \textemdash & 4 M7 Lassitude   \textemdash \\
5 G4 Tension  \textemdash & 6 B10 Hallucinations  & 4 M8 Inability to feel   \textemdash \\
3 G5 Mannerisms and posturing  \textemdash & 6 B11 Unusual thought content  & 5 M9 Pessimistic thoughts   \textemdash \\
6 G6 Depression & 5 B12 Bizarre behavior  & 5 M10 Suicidal thoughts  \textemdash \\
\hline
\end{tabular}
\caption{Symptoms included in the BPRS, PANSS and MADRS scales. The format within each cell of this table is $[$N C D$]$ where N is the maximum assigned rating observed within the data studied , C is the code name for each symptom, and D is it's descriptive name. Symptoms that were not found to be predictable by any phoneme have been marked with \textbf{--} .
Note that the allowed ranges are from 1-7 for BPRS (B*), 0-6 for
MADRS (M*) and 1-7 for PANSS (P*,N*,G*).}
\label{tab:allscales}
\end{table*}

\section{Learning voice-rating relationships} \label{sec:learning}
\subsection{Feature representation}
The conventional approach to representing speech signals is  to use Mel-frequency Cepstral Coefficients (MFCC), which describe the smoothed envelope of the speech spectrum against a perceptually-warped frequency axis. Measurements are typically taken on short 25ms analysis windows, 100 times per second. In recent studies \cite{singhiwbfsubglottal2016} we have shown that MFCCs are ineffective at capturing the contributions of key physical attributes of the speaker, namely sub-glottal resonances, and have proposed an alternate characterization of the signal based on high-resolution autoregressive (AR) modeling that shows superior performance in the extraction of speakers' physical attributes from the speech signal. In this work we use the AR-model-based features proposed in \cite{singhiwbfsubglottal2016} to represent the speech signal. The speech signal is upsampled and a 128-order AR model is estimated from it using Burg's maximum entropy method \cite{burg1968new}. A 64-point log spectrum spanning a uniformly spaced range of frequencies from 20Hz-6400Hz is obtained from the AR model. These feature vectors are subsequently used for all our analysis.
\vspace{-0.1in}
\subsection{Modeling the relationships between articulatory-phonetic units and symptoms}
We segment the speech into its constituent phonemes using an automatic speech recognition system modified explicitly for this task to yield high-accuracy segmentation. The rating of psychiatric symptoms is not expected to be linearly related to acoustic features, and hence linear regression models, and features such as correlations and $R^2$ values, that capture linear relationships between predictor and dependent variables, are unsuitable for our purpose. Instead, for each symptom, we train a non-parametric regression model to predict the rating from the acoustic features. This results in as many rating predictors for each symptom as there are phonemes in the ensemble. We quantify the relationship between the acoustic features and the symptom rating through the {\em correlation} between the predictions made by the model and the symptom rating assigned by the psychiatrist. For our experiments we use Random Forest (RF) regression \cite{liaw2002classification}, which has been shown to be effective for such studies \cite{schotz2005automatic}. In our experiments RF regression was observed to result in more accurate predictions (as measured by absolute error between the true and predicted symptom ratings) and better correlations than other models such as support-vector regression, K-nearest-neighbor models, Kernel regression etc.
\vspace{-0.1in}
\subsection{Phoneme selection and statistical significance}
\vspace{-0.1in}
Any phoneme in which the correlation between the predicted and true values of a symptom exceeds 0.2 is deemed to be predictive of that symptom. To eliminate incidental correlations, a statistical significance test is conducted to further restrict the selection.
The statistical significance of the computed correlations between predicted and actual values is determined using a $t$-test \cite{lowry2013concepts}. We note that our data actually occur in groups -- predictions from all instances of a phoneme by a given speaker may be expected to cluster together more closely than predictions from other speakers. Thus, the data do not exactly correspond to the assumptions made by the $t$-test, and the $P$ values reported by the $t$-test may be optimistic. In order to compensate for this, we use $P$ value thresholds of 0.001 and 0.0001 to report results.  Thus, all reported correlations, even if low, have high confidence.
\vspace{-0.1in}
\section{Experimental results} \label{sec:expts}
\vspace{-0.1in}
Our experimental data comprise speech recordings from 16 patient-doctor interactions, each an hour or longer in duration. The patients were not heavily accented and spoke variations of North American English. Nevertheless, since we did not normalize for accents, and it is known that vowels are relatively more influenced by accent variations, we decided to exclude vowels from our current study. The data were hand-transcribed carefully, with all speech and non-speech events, including breaths, coughs, throat-clearing etc. marked. The CMU sphinx-3 continuous speech ASR system, trained on North American English, was used to accurately extract phonemes from the conversations. For maximum accuracy, the system was acoustically adapted to each patient's speech separately. Before proceeding with the experiments, the resulting phoneme segmentations were manually checked and found to be extremely accurate. 
\vspace{-0.1in}
\begin{figure*}[ht]
\centering
\includegraphics[width=0.6\textwidth]{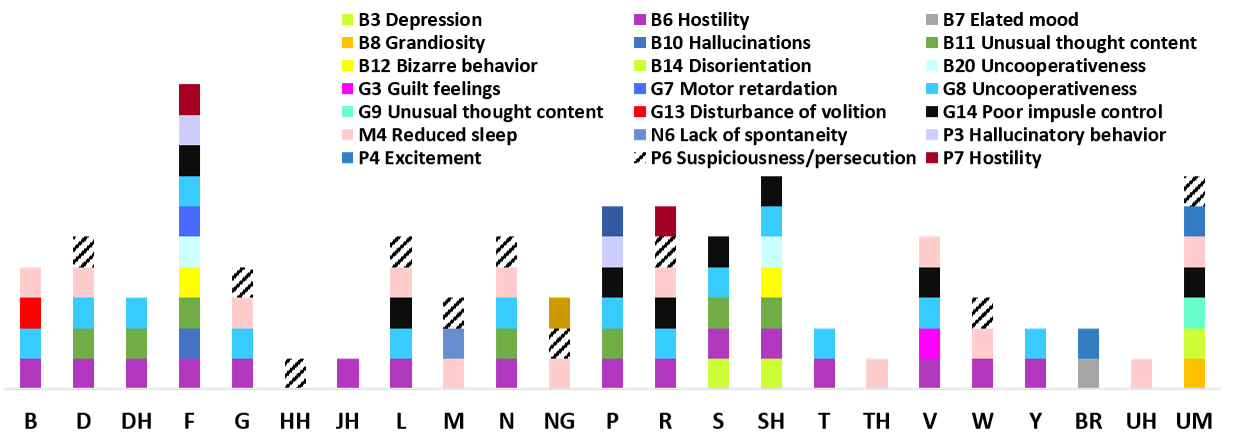}
\vspace{-0.1in}
\caption{Symptoms predicted by different consonants and the filler sounds BR: Breath, UH: uuuh.. and UM: uuum...}
\label{fig:symptomspredicted}
\end{figure*} 
\begin{figure}[ht]
\centering
\includegraphics[width=0.38\textwidth]{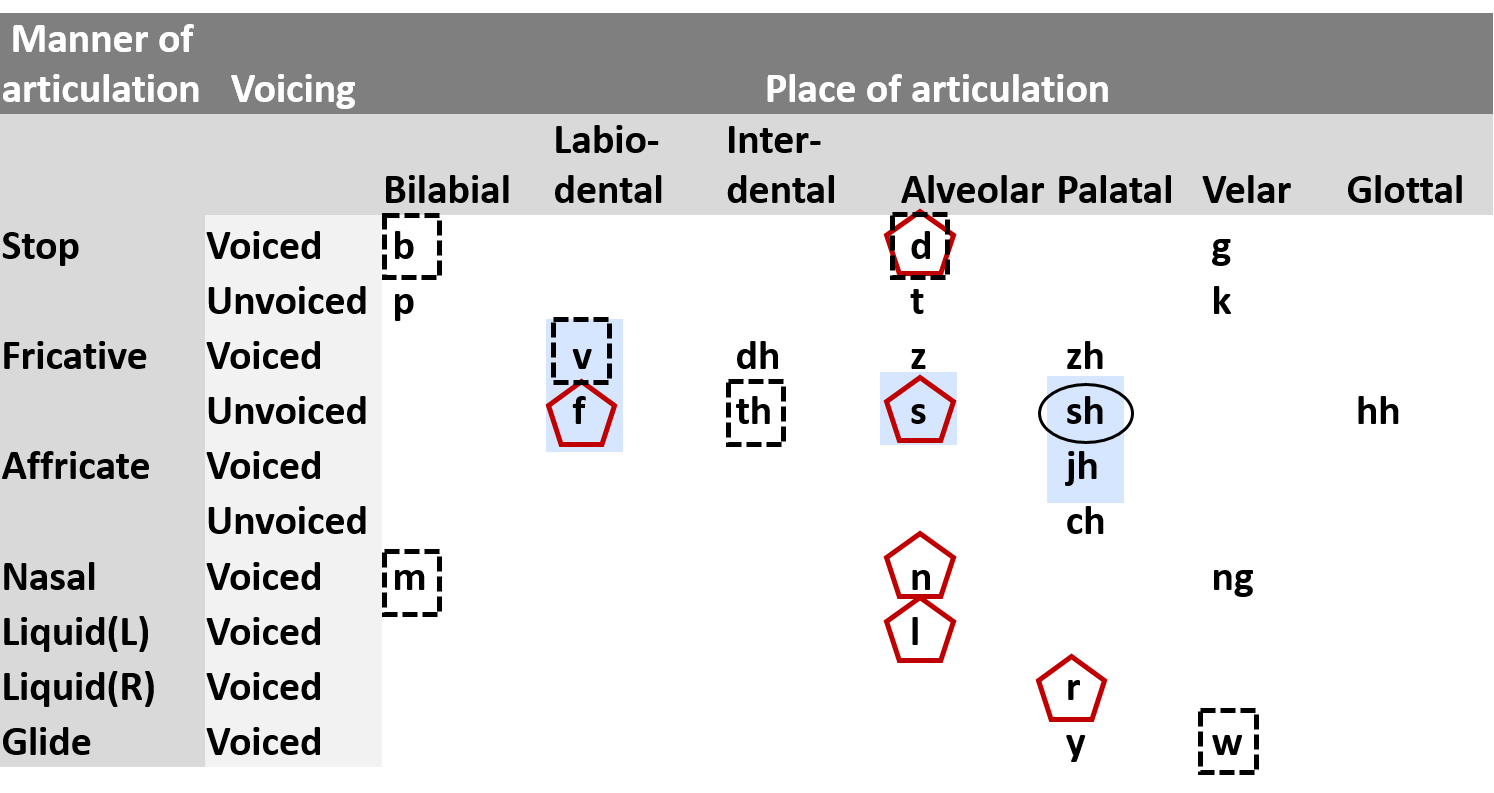}
\caption{Most predictive consonants ($P$-value$<$0.0001) for 4 symptoms: BPRS \textbf{B6} Hostility/ Animosity/Contempt (shaded blue), PANSS \textbf{G14} Poor Impulse Control (within ellipse), PANSS \textbf{G8} Uncooperativeness (within pentagon), MADRS \textbf{M4} Reduced Sleep (within dotted rectangle) }
\label{fig:consonantschart}
\end{figure}
Each experiment comprised a subset of independent sub-experiments, each of which trained and tested a 100-tree RF-based rating predictor for one symptom using AR-spectral features derived from all examples of a single phoneme. Thus, in each sub-experiment, all instances of the phoneme in consideration were collated across the ensemble of 16 patient recordings. This resulted in a working set of several tens to thousands of instances of each phoneme. Following this, a 16-fold cross-validation experiment was performed on the set where, in each fold, a specific RF predictor was trained on data from 15 speakers and tested on the 16$^{th}$. 
All 66 psychiatric symptoms listed in Table \ref{tab:allscales} were evaluated in this manner. The phoneme set comprised 50 phonemes, of which 24 were consonants and 10 were filler sounds such as the locutions "UM" and "UH", breath, throat-clearing etc. Since there were only 16 patients in the dataset, ratings for all symptoms did not fully span the available range of ratings prescribed in the corresponding scales. Table \ref{tab:allscales} also shows the maximum value of the rating for each symptom found in our dataset.

\noindent \textbf{Phoneme-level analysis: } 29 symptoms across the three scales did not show any correlation to the acoustic signal. These are indicated in Table \ref{tab:allscales}. 
Of the remaining predictable symptoms, Fig. \ref{fig:symptomspredicted} shows the symptoms that are most strongly predictable from each of the consonant and filler sounds listed. Phonemes for which the correlation between the predicted and true values of symptoms exceeded 0.2 with a $P$ value no greater than 0.001 were selected for this figure. Note that not all consonants were predictive of symptoms. Some consonants do not appear in this list at all. For a full list of consonants, see Fig. \ref{fig:consonantschart},  which also shows the 

\begin{table}[ht]
\begin{tabular}{|p{1.2cm}|p{6.0cm}|}
\hline
\textbf{Symptom} & \textbf{Most predictive consonant categories} \\ \hline
B6 & Labiodental, Palatal, Alveolar, Bilabial, Velar \\
B11 & Labiodental, Alveolar, Interdental \\
G8 & Labiodental, Palatal, Alveolar, Interdental \\
M4 & Interdental, Bilabial, Velar, Alveolar \\
P6  & Glottal, Alveolar, Velar \\
\hline
\end{tabular}
\caption{Symptoms predicted by different consonant-categories based on their place of articulation, in decreasing order of correlation. Only categories with $P$-values$<$0.001 are reported.}
\label{tab:categories}
\vspace{-0.05in}
\end{table}

the \textit{averaged} predictions from the dominant consonant categories that were generated from the ensemble of most predictive consonants. Here a tighter $P$-value threshold of 0.0001 was used to select the consonants.
\vspace{-0.2in}
\section{Conclusions} \label{sec:concl}
The phoneme (and articulator)-specificity in the exhibition of symptoms, can lead to interesting hypotheses: \textit{e.g.} some symptoms are highly correlated with labiodental fricative sounds. Since these are produced by significant turbulence in the vocal tract, involving more precise articulation on the part of the speaker, we can conjecture that the associated symptoms cause more changes in these maneuvers. The distribution of the predictability of symptoms over articulatory-phonetic charts, as in Fig. \ref{fig:consonantschart}, is useful in highlighting the specific patterns of these relationships. These may enable us to focus on specific physical manifestations of psychiatric symptoms in larger studies. Severl symptoms showed no acoustics-based predictability for any phoneme. There could be multiple reasons for this: a) there was no inter-annotator agreement performed for assigning the final ratings. which my question the accuracy of ratings. However, we note that the clinicians were highly skilled, and this is an unlikely cause; b) there are underlying correlations, but due to the paucity of data they did not surface in this analysis; c) there are truly no correlations between the spectral properties of the acoustic signal and these symptoms. We do not have enough data within this study to validate or disprove any of these conjectures.
\clearpage
\newpage

\bibliographystyle{IEEEtran}
\bibliography{psychiatric}

\end{document}